\documentclass[aps,pra,reprint,showpacs,superscriptaddress]{revtex4-1}

\usepackage{tikz}
\usepackage{pgfplots}
\usetikzlibrary{plotmarks}
\usepackage{float}
\usetikzlibrary{snakes,arrows,decorations.pathmorphing,backgrounds,positioning,fit,petri}
\usetikzlibrary{calc} 

\usepgfplotslibrary{external}
\usepackage{hyperref}
\hypersetup{
	bookmarksopen=true,%
	bookmarksnumbered=true,%
	colorlinks=true,%
   	linkcolor=blue,
    	citecolor=blue,
    	filecolor=blue,
    	urlcolor=blue,
	pdfstartview=FitH,%
	pdfnewwindow=true
}

\usepackage{graphicx}
\usepgflibrary{arrows}
\usepgflibrary[arrows]
\usetikzlibrary[arrows]
\usepackage{mathtools} \mathtoolsset{showonlyrefs}

\usepackage{txfonts}

\usepackage{microtype}

\newcommand{\bra}[1]{\langle#1\rvert}
\newcommand{\ket}[1]{\lvert#1\rangle}

\newcommand{\intk}{\int \dif k \,}

\makeatletter
\newcommand{\vast}{\bBigg@{2}}
\newcommand{\Vast}{\bBigg@{4}}
\makeatother

\newcommand{\mi}{\mathrm{i}}

\newcommand{\dif}{\mathrm{d}}

\begin{document}
\title{Few-photon single-atom cavity QED With input-output formalism in Fock Space}
\author{Eden Rephaeli}
\email{edenr@stanford.edu}
\affiliation{ Department of Applied Physics, Stanford University, Stanford, CA 94305}
\author{Shanhui Fan}
\email{shanhui@stanford.edu}
\affiliation{ Department of Electrical Engineering, Stanford University, Stanford, CA 94305}




\date{\today}

\begin{abstract}
The Jaynes-Cummings (JC) system, which describes the interaction between a cavity and a two-level atom, is one of the most important systems in quantum optics.  We obtain analytic solutions for the one- and two-photon transport in a waveguide side-coupled to the JC system using input-output formalism in Fock space. With these results, we discuss the conditions under which the JC system functions as a photon switch for waveguide photons in both the strong and weak coupling regimes\end{abstract}

\maketitle
\section{Introduction}
Controllable interaction between light and matter at the few-photon level is one of the main pursuits in quantum information processing \cite{Kimble2008}.  Much attention has focused on quantum two-level systems in the form of atoms \cite{Thompson1992} and quantum dots \cite{Yoshie2004,Khitrova2006,Press2007} in the optical frequency range, or Rydberg atoms \cite{Brune1996} and superconducting Josephson junctions \cite{Nakamura1999,Wallraff2004,Lang2011} in the microwave frequency range. All of these different forms of quantum two-level systems can be described by the same model; in this paper we refer to them simply as "two-level atom". The two-level atom is often placed in a cavity, which may enhance or inhibit spontaneous emission \cite{Kleppner1981}.  In the strong-coupling regime, the atom coherently exchanges energy with cavity photons \cite{Brune1996,Shen2007,Devoret2007,Srinivasan2007,Rephaeli2010,Shi2011}, as observed in vacuum Rabi-splitting \cite{Mondragon1983} of the out-coupled photon spectra \cite{Khitrova2006,Englund2010}.

We consider a system consisting of a waveguide side-coupled to a single-mode cavity containing a two-level atom.  An example of such system is shown in Fig. \ref{F:1}(a) for the case where both the waveguide and the cavity are in a photonic crystal. This geometry has been demonstrated in several recent experiments \cite{Wallraff2004,Srinivasan2007b,Dayan2008,Fink2008,Schoelkopf2008,Faraon2008b}. Theoretically, the atom-cavity system has been extensively studied with both numerical and analytic techniques. Numerically, this system has been simulated with an inherently stochastic quantum trajectory method \cite{Dalibard1992,Tian1992,Faraon2010}, and by directly solving the Master equation with a truncated number-state basis describing the photons in the cavity \cite{Brecha1999,Blais2004,Birnbaum2005,Srinivasan2007b}.
Analytically, this system can be analyzed with a coherent-state input by solving the time-evolution of the density matrix \cite{Agarwal1984,Agarwal1986,Carmichael1985}, or by using input-output formalism assuming with the weak-excitation approximation \cite{Thompson1992,Waks2006}.

We provide a fully quantum mechanical study of one- and two-photon waveguide transport through the atom-cavity system. Previously, single-photon transport was studied by solving for the real-space representation of the one-excitation eigenstate of the system \cite{Shen2009a,Bermel2006}, while two-photon transport was studied using the field theoretic LSZ formalism \cite{Shi2011}; both of these solutions are exact. Here, we re-derive the one- and two-photon S-matrices using input-output formalism \cite{Gardiner1985} in Fock space \cite{Fan2010}. our analytic results are in complete agreement with Refs. \cite{Shen2009a,Shi2011}.

While the analytic results we present in this paper are known in the literature, we believe our re-derivation is of significance. First of all, in comparison with the real-space wavefunction, and the LSZ field theoretic approach, input-output formalism is more widely used in quantum optics. Traditional use of input-output formalism, however, has largely focused on solving for properties of quantum systems with a coherent or squeezed state input. In this regard, it is of value to demonstrate how one can use the same formalism to solve for transport properties of Fock states in a geometry that is of direct experimental importance. Secondly, in the case of two-photon transport, input-output formalism in fact results in a much simpler and far more transparent derivation of the two-photon scattering matrix, compared with the LSZ technique. Finally, we note that our Fock-state approach may be generalized for more than two photons, and also multiple atoms \cite{Rephaeli2011}, thus complimenting approaches with similar capabilities that are based on coherent-state excitations \cite{Koshino2007, Koshino2008,Koshino2008b,Kocabas2012}. We therefore believe our work here can facilitate greater understanding of the transport properties of non-classical Fock states in this experimentally important geometry.

In Refs. \cite{Birnbaum2005,Faraon2008}, it was theoretically and experimentally demonstrated that the unequal energy level spacing in the JC ladder, shown in Fig. \ref{F:1}(b), gives rise to photon-blockade, where  a single photon blocks the transport of a second, identical photon, through a direct-coupled cavity. This effect is analogous to the Coulomb blockade effect in condensed matter physics,  where one electron blocks the transmission of a second electron through sufficiently small semiconductor islands due to the single electron charging energy \cite{Likharev1999}. Using our analytic results we show that when the cavity is side-coupled, the system can act as a two-photon switch in both the strong and weak coupling regimes, and discuss the connection to photon blockade in the direct-coupled cavity.

The structure of the paper is as follows.  In Section \ref{S:2} the Hamiltonian of the system is introduced, and equations of motion are derived.  In Sec. \ref{S:3} the single-photon S-matrix is calculated, and single-photon transport is discussed.  In Sec. \ref{S:4} the two-photon S-matrix and the response to a two-photon planewave are calculated.  In Sec. \ref{S:5} we discuss the two-photon switch behavior.

\section{Hamiltonian and equations of motion}
\label{S:2}
\begin{figure}
\includegraphics{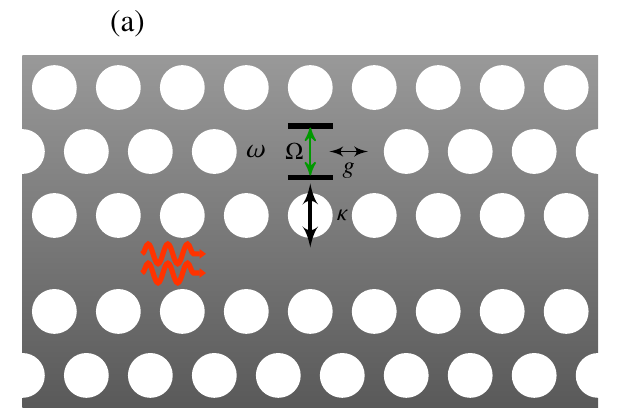}
\includegraphics{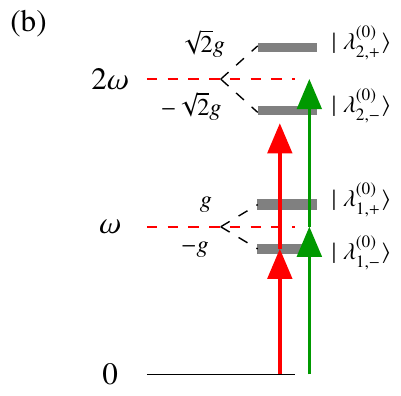}
\caption{(a) schematic of a two-level atom in a cavity side-coupled to a waveguide. (b) the energy spectrum of the JC hamiltonian ($\Omega=\omega$) in the excitation number manifolds n = 0, 1, and 2.  Arrows show one and two photon absorption for two-photon switch.  The switch is observed in reflection in the  strong-coupling regime (red arrows), and in transmission in the weak-coupling regime (green arrows).}
  \label{F:1}
\end{figure}

A cavity containing a single two-level atom is described by the Jaynes-Cummings (JC) hamiltonian \cite{Jaynes1963}: 
\begin{align}
H_{JC}=\frac{1}{2}\Omega\sigma_z+\omega c^{\dagger}c+g\left[ c^{\dagger}\sigma_-+\sigma_+c \right]\label{E:HJC},
\end{align}
where $\Omega$ is the atomic transition frequency, $\omega$ is the cavity mode frequency, and $g$ is the atom-cavity coupling rate. The atom and cavity operators satisfy $\left[\sigma_+,\sigma_-\right]=\sigma_z$ and $\left[c,c^{\dag}\right]=1$, respectively.   
We side-couple the atom-containing cavity described by Eq. \eqref{E:HJC} to a waveguide with linear dispersion relation and group velocity $v_g$ for right- and left-moving photons [Fig. \ref{F:1}(a)], with respective real-space photon creation (annihilation)  operators $a^{\dag}_R(x) [a_R(x)]$ and $a^{\dag}_L(x) [a_L(x)]$, satisfying $\left[a_R(x),a^{\dag}_R(x')\right]=\left[a_L(x),a^{\dag}_L(x')\right]=\delta(x-x')$.  We refer to this atom-cavity-waveguide system as the \textit{two-mode model}.
To solve the two-mode model, we exploit the spatial inversion symmetry of the system by decomposing the Hilbert Space into even and odd subspaces with respective photon operators $a_{e}(x)\equiv\frac{1}{\sqrt{2}}\left[a_R(x)+a_L(-x)\right]$ and $a_{o}(x)\equiv\frac{1}{\sqrt{2}}\left[a_R(x)-a_L(-x)\right]$, satisfying  $\left[a_e(x),a^{\dag}_e(x')\right]=\left[a_o(x),a^{\dag}_o(x')\right]=\delta(x-x')$.  Both even and odd subspaces feature a one-way (chiral) waveguide mode, but only the waveguide mode in the even subspace couples to the JC system.  We refer to the system in the even subspace as the \textit{one-mode model}. We will solve for the S-matrix of the one-mode model, from which the two-mode model S-matrix is then straightforwardly constructed (See Appendix A in Ref. \cite{Fan2010}).
The one-mode model hamiltonian is given by:  
\begin{align}
H_{\text{e}}=v_g\intk k\ a^{\dagger}_ka_k+V\int dk \left[ a^{\dagger}_kc+c^{\dagger}a_k\right]+H_{JC},\label{E:Htotal} 
\end{align}
where $V$ is the waveguide-cavity coupling strength. The waveguide photon operator $a_k=\frac{1}{\sqrt{2\pi}}\int dx e^{ikx}a_e(x)$ satisfies $\left[a_k,a^{\dag}_k\right]=\delta(k-k')$, where we omitted the $e$ subscript from the $k$-space photon operators.  Throughout this paper we set $v_g=1$.

Heisenberg operator equations for $ a_k(t)$, $\sigma_-(t)$ and $c(t)$ follow from Eq. \eqref{E:Htotal}:
\begin{align}
&\frac{\dif a_k(t)}{\dif t}=-\mi ka_k(t)-\mi Vc(t), \label{E:ak}
 \\
&\frac{\dif c(t)}{\dif t}=-\mi\omega c(t)-\mi V\int dka_k(t)-\mi g\sigma_-(t), \label{E:cdot1}
 \\
&\frac{\dif\sigma_-(t)}{\dif t}=-\mi\Omega\sigma_-(t)+\mi g\sigma_z(t)c(t).\label{E:sigdot1}\end{align}
In Appendix \ref{AppB} we follow the approach of ref. \cite{Fan2010}, defining input and output waveguide photon operators, leading to
\begin{align}
&a_{\text{out}}(t)=a_{\text{in}}(t)-\mi\sqrt{\kappa}\ c(t),
\label{E:aoutain}
\\&\frac{dc(t)}{dt}=\left(-\mi\omega-\frac{\kappa}{2}\right)c(t)-\mi\sqrt{\kappa}\ a_{\text{in}}(t)-\mi g\ \sigma_-(t).
\label{E:cdot2}
\end{align}
where $\kappa=2\pi V^2$.  Eqs. \eqref{E:sigdot1}, \eqref{E:aoutain} and  \eqref{E:cdot2} are the fundamental equations in the input-output formalism for this system.  Here we will solve these equations for one and two-photon scattering matrix (S-matrix).

\section{One-Photon Scattering Matrix}
\label{S:3}
The single photon S-matrix between two free single-photon states with energies $k$ and $p$ may be expressed using input and output operators as \cite{Fan2010}:
\begin{align}
_{e}\bra{p}\textbf{S}\ket{k}_{e}=\langle p^{-}\mid k^{+}\rangle=\bra{0}a_{out}(p)a^{\dag}_{in}(k)\ket{0}.
\label{E:Sone}
\end{align}
Here, $\ket{k^+}$ and $\ket{p^-}$ are scattering eigenstates that evolve in the interaction picture from free-photon states in the distant past and future, respectively \cite{Fan2010}. Making use of  
\begin{align}
a_{\text{out}}(k)=a_{\text{in}}(k)-\mi \sqrt{\kappa}\ c(k),\label{E:aoutaink}
\end{align}
the Fourier transform of Eq. (\ref{E:aoutain}), we rewrite Eq. \eqref{E:Sone} using input-field operators:
\begin{align}
\langle p^{-}\mid k^{+}\rangle=\delta(k-p)-\mi\sqrt{\kappa}\bra{0}c(p)\ket{k^{+}}.
\label{E:SoneIn}
\end{align}
To calculate the matrix element $ \langle0\mid c(p)\mid k^{+}\rangle$, we derive an equation of motion for its time-domain counterpart $\bra{0}c(t)\ket{k^+}$ using Eq. \eqref{E:cdot2}, resulting in
\begin{align}
&\frac{d}{dt}\bra{0}c(t)\ket{k^{+}}=\left(-\mi\omega-\frac{\kappa}{2}\right)\bra{0}c(t)\ket{k^{+}}-\mi\sqrt{\kappa}\bra{0}a_{\text{in}}(t)\ket{k^{+}}\\&-\mi g\bra{0}\sigma_-(t)\ket{k^{+}},
\label{E:cdot3}
\end{align}
where $\bra{0}a_{\text{in}}(t)\ket{k^+}=\frac{1}{\sqrt{2\pi}}e^{-\mi kt}$, and from Eq. \eqref{E:sigdot1} $\bra{0}\sigma_-(t)\ket{k^{+}}$ satisfies
\begin{align}
&\frac{d}{dt}\bra{0}\sigma_-(t)\ket{k^{+}}=-\mi\Omega\bra{0}\sigma_-(t)\ket{k^{+}}-\mi g\bra{0}c(t)\ket{k^{+}}.\label{E:sigdot2}
\end{align}
Here, we have used the fact that $\sigma_z\ket{0}=-\ket{0}$.  Eq. \eqref{E:sigdot2} can be obtained alternatively using the weak excitation approximation by setting $\sigma_z\to-1$ in Eq. \eqref{E:sigdot1}.  Thus, the weak excitation approximation is in fact justified when treating single-photon transport.  However, in the single photon calculation, the approach here can also be used to directly calculate the excitation of the atom.

The solutions to Eqs. \eqref{E:cdot3} and \eqref{E:sigdot2} yield:
\begin{align}
&\bra{0}c(p)\ket{k^+}=\frac{ \sqrt{\kappa}\left(k-\Omega\right)}{\left(k-\omega+\mi\frac{\kappa}{2}\right)\left(k-\Omega\right)-g^2}\delta(k-p)\\&\equiv s^{(c)}_k\delta(k-p)
\label{E:sck}\\
&\bra{0} \sigma_-(p)\ket{k^{+}}= \frac{\sqrt{\kappa}g}{\left(k-\omega+\mi\frac{\kappa}{2}\right)\left(k-\Omega\right)-g^2}\delta(k-p)\\&\equiv s^{(a)}_k\delta(k-p)
\label{E:sak}
\\&\langle p^{-}\ket{k^{+}}=\frac{\left(k-\omega-\mi\frac{\kappa}{2}\right)\left(k-\Omega\right)-g^2}{\left(k-\omega+\mi\frac{\kappa}{2}\right)\left(k-\Omega\right)-g^2}\delta(k-p)\\&\equiv t_k\delta(k-p)
\label{E:Skp}
\end{align}
where $s^{(c)}(k)$, $s^{(a)}(k)$ and $t_k$ are the cavity and atom excitation amplitudes, and the one-mode transmission coefficient, respectively.  We note that the poles of all three functions are $\lambda_{1,\pm}=\frac{\omega+\Omega-\mi\frac{\kappa}{2}}{2}\pm \sqrt{\left(\frac{\omega-\Omega-\mi\frac{\kappa}{2}}{2}\right)^2+g^2}$. In the limit $\kappa\to 0$ we have $\lambda_{1,\pm}$ approaching the one-excitation eigenvalues $\lambda^{(0)}_{1,\pm}$ of the JC Hamiltonian.

Having derived the scattering matrix for the one-mode model above, we can straightforwardly obtain the transmission and reflection amplitudes for the two-mode model:
\begin{align}
&_{R}\bra{p}\textbf{S}\ket{k}_R=\bar{t}_k\delta(k-p),
\\&_{L}\bra{p}\textbf{S}\ket{k}_R=\bar{r}_k\delta(k+p),
\end{align}
where $\bar{t}_k=\frac{1}{2}\left(t_k+1\right)$, $\bar{r}_k=\frac{1}{2}\left(t_k-1\right)$ and $\ket{k}_{R,L}=\frac{1}{\sqrt{2\pi}}\int dx e^{ikx}a^{\dag}_{R,L}(x)\ket{0}$.  Below, we provide a brief discussion of the properties of the single-photon transport for the two-mode model, while only highlighting those aspects that are relevant for the discussions of the two-photon properties.  A detailed discussion of single-photon transport in this system can be found in Ref. \cite{Shen2009a}.  
\begin{figure}
\includegraphics{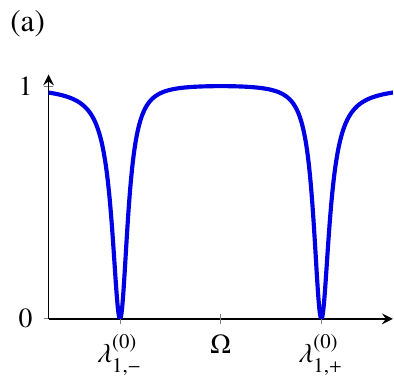}
\includegraphics{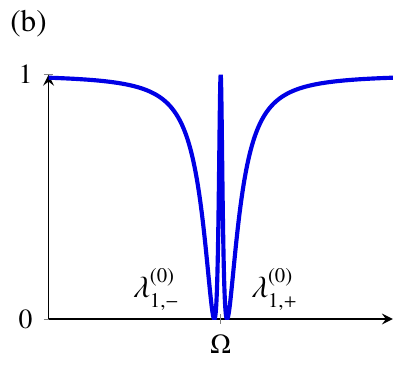}
\caption{Single-photon transmission $|\bar{t}_k|^2$ vs. incoming photon angular frequency in the case of tuned atom and cavity ($\omega=\Omega$).  (a) strong-coupling regime ($g>\kappa$). (b) weak-coupling regime ($g<\kappa$).}
  \label{F:2}
\end{figure}

In Fig. \ref{F:2}, the transmission spectrum for the case of atom and cavity on-resonance $(\omega=\Omega)$ is plotted.  We plot both the weak coupling regime with $g<\kappa$ [Fig. \ref{F:2}(a)], and the strong coupling regime with $g>\kappa$ [Fig. \ref{F:2}(b)].  In both regimes,  three extrema are present in the spectrum. Two transmission minima where the incident photon is completely reflected occur at $k=\lambda^{(0)}_{1,\pm}=\omega\pm g$, the single-excitation energy eigenvalue of the JC hamiltonian.  At these frequencies, $s_{k=\lambda^{(0)}_{1,\pm}}^{(c)}=\pm s_{k=\lambda^{(0)}_{1,\pm}}^{(a)}$, the cavity and atom excitations have equal amplitudes and either equal or opposite phase. .  The transmission spectra have a maximum at $k=\Omega$, where the incident photon is fully transmitted.  At this frequency, the cavity excitation amplitude is zero ($s_{k=\Omega}^{(c)}=0$), while the atomic excitation amplitude is maximal $s_{k=\Omega}^{(a)}=-\sqrt{\kappa}/g$.  This is the dipole-induced transparency effect pointed out in Ref. \cite{Waks2006}. In the strong coupling regime [Fig. \ref{F:2}(a)], the spectral width of each transmission dip is $\kappa/2$, independent of the value of $g$, while in the weak-coupling regime the transmission spectrum displays the electromagnetically-induced transparency (EIT) type of feature, with a narrow central transmission peak whose width is proportional to $g$, as shown in Fig. \ref{F:2}(b).   
\section{Two-Photon Scattering Matrix}
 \label{S:4}
In the one-mode model, the two-photon S-matrix $_{ee}\bra{p_1,p_2}\textbf{S}_{ee}\ket{k_1,k_2}_{ee}=\bra{p_1p_2^-}k_1k_2^+\rangle$, connecting an incoming two-photon state with photon energies $k_1$ and $k_2$ to an outgoing two-photon state with photon energies $p_1$ and $p_2$, may be written as \cite{Fan2010}:
\begin{align}
&\bra{0} a_{out}(p_2) a_{out}(p_1) a^{\dag}_{in}(k_2) a^{\dag}_{in}(k_1) \ket{0}\\&=\int dp\bra{p_1^-}p^+\rangle\bra{p^+} a_{out}(p_2)  a^{\dag}_{in}(k_2) a^{\dag}_{in}(k_1) \ket{0}\\&=t_{p_1}\bra{p^{+}_1} a_{out}(p_2)  a^{\dag}_{in}(k_2) a^{\dag}_{in}(k_1) \ket{0}
\label{E:Stwo}
\end{align}
where we have inserted a complete single-photon basis above, and made use of the one-photon S-matrix from Sec. \ref{S:3}.  Substituting Eq. \eqref{E:aoutaink} we obtain:
\begin{align}
&\bra{p_1p_2^-}k_1k_2^+\rangle=\\&t_{p_1}\bra{p^{+}_1}a_{in}(p_2)\ket{k_1k_2^{+}}-\mi\sqrt{\kappa}\bra{p_1^+}c(p_2)\ket{k_1k_2^{+}}
\label{E:StwoIn}
\end{align}
In order to calculate $\bra{p^{+}_1}c(p_2)\ket{k_1k_2^+}$, we obtain a differential equations for its time-domain counterpart $\bra{p^+_1}c(t)\ket{k_1k_2^+}$  using  Eq. \eqref{E:cdot1}:
\begin{align}
&\frac{d}{dt}\bra{p^{+}_1}c(t)\ket{k_1k^{+}_2}=-\mi\left(\omega-\mi\frac{\kappa}{2}\right)\bra{p^{+}_1}c(t)\ket{k_1k^{+}_2}\\&-\mi\sqrt{\kappa}\bra{p^{+}_1}a_{in}(t)\ket{k_1k^{+}_2}-\mi g\bra{p^{+}_1}\sigma_-(t)\ket{k_1k^{+}_2}
\label{E:sig1d}\\
\end{align}
where $\bra{p^{+}_1}\sigma_-(t)\ket{k_1k^{+}_2}$ satisfies
\begin{align}
&\frac{d}{dt}\bra{p^{+}_1}\sigma_-(t)\ket{k_1k^{+}_2}=-\mi\Omega\bra{p^{+}_1}\sigma_-(t)\ket{k_1k^{+}_2}\\&+\mi g\bra{p^{+}_1}\sigma_z(t)c(t)\ket{k_1k^{+}_2}\label{E:sig2d}
\end{align}
as can be obtained using Eq. \eqref{E:sigdot1}. We must now solve for the matrix element $\bra{p^{+}_1}\sigma_z(t)c(t)\ket{k_1k^{+}_2}$ in Eq. \eqref{E:sig2d}, which may be rewritten using the identity $\sigma_z=2\sigma_+\sigma_--1$ as: 
\begin{align}
&\bra{p^{+}_1}\sigma_z(t)c(t)\ket{k_1k^{+}_2}=2\bra{p^{+}_1}\sigma_+(t)\ket{0}\bra{0}\sigma_-(t)c(t)\ket{k_1k^{+}_2}\\&-\bra{p^{+}_1}c(t)\ket{k_1k^{+}_2}
\label{E:sigmazIden}
\end{align}
Examining Eq. \eqref{E:sigmazIden}, we note that the weak-excitation approximation ($\sigma_z\to -1$) certainly does not hold. The first term on the right-hand side of Eq. \eqref{E:sigmazIden}, which would be completely absent in the weak-excitation limit, is in fact responsible for two-photon resonances in the S-matrix.
To proceed, we generate an equation for the matrix element $\bra{0}\sigma_-(t)c(t)\ket{k_1k^{+}_2}$ 
\begin{align}
&\frac{d}{dt}\bra{0}\sigma_-(t)c(t)\ket{k_1k^{+}_2}=-\mi\left(\omega+\Omega-\mi\frac{\kappa}{2}\right)\bra{0}\sigma_-(t)c(t)\ket{k_1k^{+}_2}\\&-\mi\sqrt{\kappa}\bra{0}\sigma_-(t)a_{in}(t)\ket{k_1k^{+}_2}-\mi g\bra{0}c^2(t)\ket{k_1k^{+}_2}\label{E:sigcdot}
\end{align}
where $\bra{0}c^2(t)\ket{k_1k_2^+}$ satisfies
\begin{align}
&\frac{d}{dt}\bra{0}c^{2}(t)\ket{k_1k^{+}_2}=-2\mi\left(\omega-\mi\frac{\kappa}{2}\right)\bra{0}c^{2}(t)\ket{k_1k^{+}_2}
\\&-2\mi\sqrt{\kappa}\bra{0}c(t)a_{in}(t)\ket{k_1k^{+}_2}-2\mi g\bra{0}\sigma_-(t)c(t)\ket{k_1k^{+}_2}.
\label{E:csqdot}
\end{align}
In deriving Eq. \eqref{E:csqdot} we have used the identity $\left[c(t),a_{\text{in}}(t)\right]=0$ (see Appendix A in Ref. \cite{Rephaeli2011} for a similar proof). 
We now have a closed set of ordinary differential equations for the various matrix elements involved. Eqs. \eqref{E:sigcdot} and \eqref{E:csqdot} may then be solved for $\bra{0}\sigma_-(t)c(t)\ket{k_1k^{+}_2}$.  Using this solution, we solve Eqs. \eqref{E:sig1d} and \eqref{E:sig2d}, and obtain the two-photon S-matrix:
\begin{align}
&\bra{p_1p_2^-}k_1k_2^+\rangle=t_{p_1}t_{p_2}\left[\delta(k_1-p_1)\delta(k_2-p_2)+\delta(k_1-p_2)\delta(k_2-p_1)\right]\\
&+B\delta(E_o-E_i),
\label{E:sp1p2k1k2}
\end{align}
where 
\begin{align}
&B=i\frac{\sqrt{\kappa}g}{\pi}s^{(a)}_{p_1}s^{(a)}_{p_2}
\frac{2g\left[s^{(c)}_{k_1}+s^{(c)}_{k_2}\right]+\left(E_i-2\omega+i\kappa\right)\left[s^{(a)}_{k_1}+s^{(a)}_{k_2}\right]}{\left(E_i-\lambda_{2,+}\right)\left(E_i-\lambda_{2,-}\right)}
\label{E:B}
\end{align}
is the fluorescent term---the source of photon correlation effects,
and $E_i=k_1+k_2$, $E_o=p_1+p_2$ are the total energy of the incident and outgoing photons, respectively.  The above S-matrix is identical to $S_{p_1p_2k_1k_2}$ in Eq. (6) of ref. \cite{Shi2011}, which was obtained using the field-theoretic LSZ method. The present derivation is more elementary.  Also,  in the present derivation, the S-matrix' composition in terms of one-photon excitation amplitudes and the role of the two-photon poles, $ \lambda_{2,\pm}=\frac{\Omega+3\omega-\mi \frac{3\kappa}{2}}{2}\pm\sqrt{\left(\frac{\Omega-\omega+\mi\frac{\kappa}{2}}{2}\right)^2+2g^2}$, is transparent.  

Starting from the S-matrix for the one-mode model in Eq. \eqref{E:sp1p2k1k2}, we obtain the scattering amplitudes in the two-mode model between planewave states corresponding to right- and left-propagating photons following the procedure in Ref. \cite{Shen2007}:
\begin{align}
&_{RR}\bra{p_1p_2}S\ket{k_1k_2}_{RR}=\bar{t}_{k_1}\bar{t}_{k_2}\left[\delta(k_1-p_1)\delta(k_2-p_2)+\delta(k_1-p_2)\delta(k_2-p_1)\right]
\\&+\frac{1}{4}B\delta(E_i-E_o)
\\&_{LL}\bra{p_1p_2}S\ket{k_1k_2}_{RR}=\bar{r}_{k_1}\bar{r}_{k_2}\left[\delta(k_1+p_1)\delta(k_2+p_2)+\delta(k_1+p_2)\delta(k_2+p_1)\right]
\\&+\frac{1}{4}B\delta(E_i-E_o)
\\&_{RL}\bra{p_1p_2}S\ket{k_1k_2}_{RR}=\bar{t}_{k_1}\bar{r}_{k_2}\left[\delta(k_1-p_1)\delta(k_2+p_2)+\delta(k_1-p_2)\delta(k_2+p_1)\right]
\\&+\frac{1}{4}B\delta(E_i-E_o)
\end{align}

We consider an incident two-photon planewave state $\ket{k_1k_2}_{RR}$, comprised of two right-going photons with individual energies $k_1$ and $k_2$, as described by:
\begin{align}
\ket{k_1k_2}_{RR}=\int dx_1dx_2S_{k_1,k_2}(x_1,x_2)\frac{1}{\sqrt{2}}a^{\dag}_{R}(x_1)a^{\dag}_R(x_2)\ket{0},
\end{align}
where $S_{k_1,k_2}(x_1,x_2)=\frac{1}{\sqrt{2}2\pi}\left[e^{ik_1x_1}e^{ik_2x_2}+e^{ik_1x_2}e^{ik_2x_1}\right]$ is a symmetrized two-photon planewave.
The resulting outgoing state $\ket{\phi}$, calculated in Appendix \ref{AppC}, consists of three two-photon states
\begin{align}
\ket{\phi}=\ket{\phi}_{RR}+\ket{\phi}_{LL}+\ket{\phi}_{RL},
\end{align}
describing two right-moving, two-left moving, and a right and left moving photons, respectively, given by:
\begin{align}
&\ket{\phi}_{RR}=\int dx_1dx_2\Bigg\{\bar{t}_{k_1}\bar{t}_{k_2}S_{k_1,k_2}(x_1,x_2)+H(x_1,x_2)\Bigg\}\times
\\&\frac{1}{\sqrt{2}}a^{\dag}_R(x_1)a^{\dag}_R(x_2)\ket{0}\label{E:phirr},
\\&\ket{\phi}_{LL}=\int dx_1dx_2\Bigg\{\bar{r}_{k_1}\bar{r}_{k_2}S_{k_1,k_2}(x_1,x_2)+H(x_1,x_2)\Bigg\}\times
\\&\frac{1}{\sqrt{2}}a^{\dag}_L(-x_1)a^{\dag}_L(-x_2)\ket{0}\label{E:phill},
\\&\ket{\phi}_{RL}=\int dx_1dx_2\Bigg\{\frac{1}{2\pi}\left[\bar{t}_{k_1}\bar{r}_{k_2}e^{2i\Delta_i x}+\bar{t}_{k_2}\bar{r}_{k_1}e^{-2i\Delta_i x}\right]+\sqrt{2} H(x_1,x_2)\Bigg\}\times
\\&a^{\dag}_R(x_1)a^{\dag}_L(-x_2)\ket{0}\label{E:phirl},
\end{align}
where 
\begin{align}
&H(x_1,x_2)\equiv\frac{\mi g^2\kappa}{4\sqrt{2}}\frac{F(k_1,k_2)e^{\mi E_i x_c}}{\left(\lambda_{1,+}-\lambda_{1,-}\right)\left(E_i-\lambda_{1,-}-\lambda_{1,+}\right)}\times
\\&\Bigg[\frac{e^{i\left(E_i-2\lambda_{1,-}\right)|x|/2}}{\left(E_i-2\lambda_{1,-}\right)}
-\frac{e^{i\left(E_i-2\lambda_{1,+}\right) |x|/2}}{\left(E_i-2\lambda_{1,+}\right)}\Bigg].\end{align}
Above, we have defined the two-photon center-of-mass $x_c\equiv\frac{x_1+x_2}{2}$, and spatial separation $x\equiv x_1-x_2$ coordinates; we also define $\Delta_i=\frac{k_1-k_2}{2}$, $\Delta_o=\frac{p_1-p_2}{2}$ and
\begin{align}
&F(k_1,k_2)\equiv\mi \frac{\sqrt{\kappa}g}{\pi}
\frac{2g\left[s^{(c)}_{k_1}+s^{(c)}_{k_2}\right]+\left(E_i-2\omega+\mi\kappa\right)\left[s^{(a)}_{k_1}+s^{(a)}_{k_2}\right]}{\left(E_i-\lambda_{2,+}\right)\left(E_i-\lambda_{2,-}\right)}.\label{E:F}
\end{align}

Photon statistics of the transmitted and reflected two-photon states is studied through the second-order coherence function $g^{(2)}(\tau)=G^{(2)}(\tau)/|G^{(1)}(0)|^2$, where $G^{(1)}(\tau)=\ _{FF}\bra{\phi}a^{\dag}_F(y+\tau)a_F(y)\ket{\phi}_{FF}$, $G^{(2)}(\tau)=_{FF}\bra{\phi}a^{\dag}_F(y)a^{\dag}_F(y+\tau)a_F(y+\tau)a_F(y)\ket{\phi}_{FF}$, and $F=R,L$.  Here we calculate $g^{(2)}$ using the expression $g^{(2)}(\tau)=G^{(2)}(\tau)/G^{(2)}(\tau\to \infty)$, which is a consequence of $\lim_{\tau\to\infty}g^{(2)}(\tau)=1$.  $g^{(2)}(\tau)$ is accessible experimentally through two-photon coincidence counting.  With the help of $g^{(2)}(\tau)$, we now discuss how the JC system can function as a two-photon switch.

\begin{figure}
\includegraphics{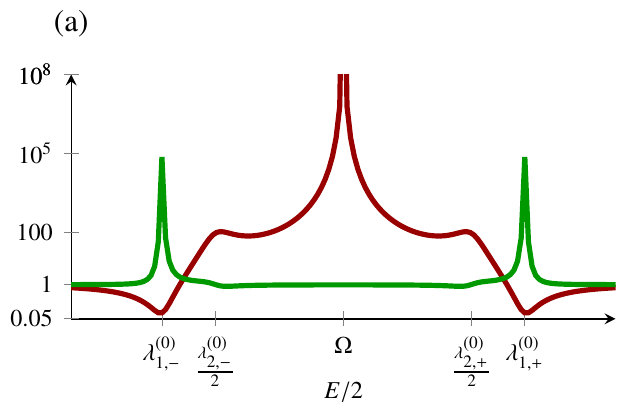}
\includegraphics{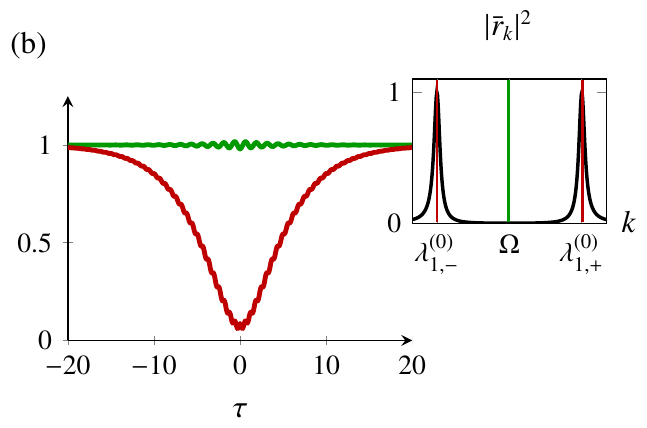}
\caption{ Strong coupling regime ($g=\sqrt{5}\kappa$). (a) Reflected $g^{(2)}(0)$ vs. $E/2$ in red curve; Transmitted $g^{(2)}(0)$ vs. $E/2$ in green curve. (b) Reflected two-photon $g^{(2)}(\tau)$  for $E/2=\lambda^{(0)}_{1,\pm}$ in red curve; transmitted two-photon $g^{(2)}(\tau)$ for $E/2=\Omega$ is green curve. Inset: Single photon reflection spectrum $|r_k|^2$}
  \label{F:4}
\end{figure}

\section{Two-photon switch}
\label{S:5}

Based on the analytic results presented above, we provide a discussion of two scenarios where the system can behave as a two-photon switch, for energy degenerate photons. Our discussion of the strong-coupling regime mirrors that of ref. \cite{Shi2011}.  As a point of departure, here we also discuss a photon-switch in the weak-coupling regime, thus providing a unified discussion of both regimes. 

\subsection{Strong-coupling regime}
The two-photon switch in the strong-coupling regime is a consequence of 
the anharmonicity of the JC ladder \cite{Birnbaum2005}: As discussed in Sec. \ref{S:3}, one incident photon tuned to the single-excitation eigenstate of the JC system, $k=\lambda^{(0)}_{1,\pm}$, is fully reflected. However, the energy of two such incident photons $2k=E_i=2\lambda^{(0)}_{1,\pm}$, does not coincide with a two-excitation eigenstate of the JC system, as illustrated by red arrows in Fig. \ref{F:1}(b).  Consequently, the two photons are prevented from simultaneously entering the cavity. Since reflection in the side-coupled cavity considered here arises entirely from cavity field decay, sub-poissonian statistics  $g^{(2)}(0)<<1$ is observed for the two reflected photons, as shown in Fig. \ref{F:4}(a). Additionally, $g^{(2)}(\tau)$ of the two reflected photons exhibits strong anti-bunching and sub-poissonian statistics for all time intervals $\tau$, as shown in Fig. \ref{F:4}(b). Here, a sufficiently large coupling rate $g$ is needed to ensure that the two-photon energy does not overlap with the two-excitation energy eigenstates, which are broadened due to waveguide coupling. We note that the photon switch behavior in reflection described above in the side-coupled cavity is identical to the photon-blockade reported in transmission through a direct-coupled cavity \cite{Birnbaum2005,Faraon2008}. 

In addition to the photon switching behavior in reflection for $E_i/2=\lambda^{(0)}_{1,\pm}$, the $g^{(2)}(0)$ spectrum also exhibits other interesting features. In reflection, $g^{(2)}(0)$ peaks at $E_i/2=\Omega$, exhibiting super-poissonian statistics [Fig. \ref{F:1}(a)].  This behavior is attributed to the vanishing single-photon reflection at $k=\Omega$, since $g^{(2)}(\tau)$ is normalized with respect to $G^{(1)}(0)$, which vanishes when single-photon reflection vanishes. We note that in the direct-coupled cavity system, the super-poissonian statistics shown here in two-photon reflection at $E_i/2=\Omega$ occurs in transmission instead \cite{Faraon2008,Kubanek2008}, and is named 'photon-induced tunneling' \cite{Faraon2008}. In transmission, $g^{(2)}(0)$ peaks at $E_i/2=\lambda^{(0)}_{1,\pm}$, displaying super-poissonian statistics [Fig. \ref{F:1}(a)], which is attributed to the vanishing single-photon transmission at $k_1=\lambda^{(0)}_{1,\pm}$.  Finally, in transmission, $g^{(2)}(0)\approx 1$ for all time intervals $\tau$ at $E_i/2=\Omega$.  At this frequency, the atom excitation scales as $1/g$, and is therefore weakly excited under strong-coupling, resulting in poissonian statistics for the transmitted light [Fig. \ref{F:1}(a)]. 

\subsection{Weak-coupling regime}
\begin{figure}
\includegraphics{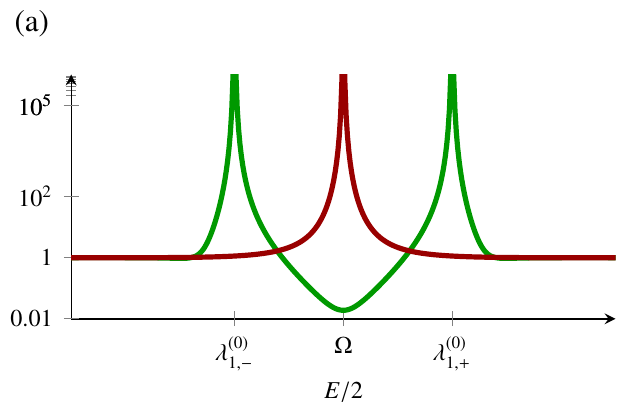}
\includegraphics{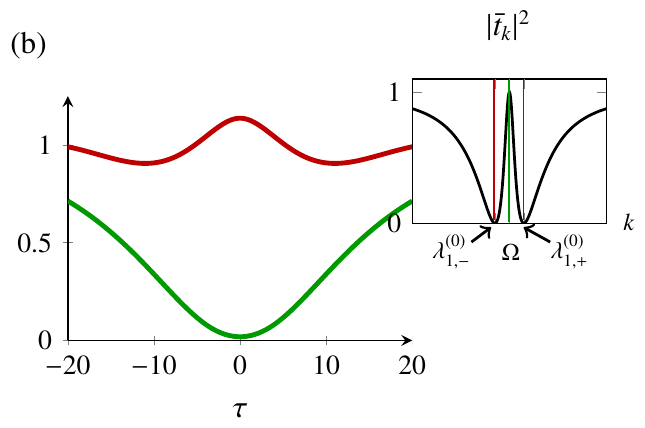}
\caption{Weak coupling regime ($g=\frac{\kappa}{\sqrt{5}}$). (a) Reflected $g^{(2)}(0)$ vs. $E/2$ in red curve; Transmitted $g^{(2)}(0)$ vs. $E/2$ in green curve. (b) Reflected two-photon $g^{(2)}(\tau)$  for $E/2=\lambda^{(0)}_{1,s}$ in red curve; transmitted two-photon $g^{(2)}(\tau)$ for $E/2=\Omega$ is green curve. Inset: Single photon transmission.}
  \label{F:5}
\end{figure}

A two-photon switch in the weak-coupling regime is also possible in this system, without requiring a nonlinear cavity \cite{Imamoglu1997}.  Here, the photon switch is created due to the intrinsic non-linearity of the atom \cite{Aoki2009,Rosenblum2011}. As discussed in Sec. \ref{S:3}, one incident photon tuned to  $k=\Omega$, is completely transmitted.  However, when two such photons, each with an energy $E_i=2\Omega$, are incident, sub-poissonian statistics $g^{(2)}(0)<<1$ is observed for the two transmitted photons, as shown in Fig. \ref{F:5}(a).  $g^{(2)}(\tau)$ of the two transmitted photons is anti-bunched, and displays sub-poissonian statistics for all time intervals $\tau$, shown in Fig. \ref{F:5}(b).  Thus, the presence of the atom in the cavity results in complete single-photon transmission (dipole-induced transparency), but the inability of the atom to absorb more than one photon at a time leads to anti-bunching and sub-poissonian statistics in two-photon transmission. We note that a similar atom-induced photon switch occurs when the atom is directly coupled to a waveguide \cite{Shen2007}. There, a single-photon is completely reflected, and anti-bunching is observed in two-photon reflection $g^{(2)}(0)=0$, while bunching is observed in two-photon transmission.

 Moreover, the two-photon reflection at $E_i/2=\Omega$ displays super-poissonian statistics as a result of the vanishing single-photon reflection at $k=\Omega$.    Finally, as is the case of strong-coupling, super-poissonian statistics is observed in two-photon transmission at $E_i/2=\lambda^{(0)}_{1,\pm}$.

 \section{Final remark and Conclusion}
As a final remark, we note that in practice, in addition to coupling to the cavity which is in turn coupled to the waveguide, the atom also couples to non guided modes, leading to loss.  This loss may be accounted for by making the replacement $\Omega\to \Omega-\mi\gamma/2$ where $\gamma$ is the coupling rate into non-guided modes \cite{Rephaeli2012}.  For the case of the D2 transition of a Caesium atom placed in a Fabry-Perot cavity, this loss can be on the order of the cavity linewidth \cite{Birnbaum2005}.  In contrast, for the case of a self-assembled InAs quantum dot in a photonic crystal cavity, the atom loss is roughly two orders of magnitude smaller than the cavity linewidth \cite{Faraon2008}, and therefore does not lead to a qualitative difference in the transport properties with respect to the non-lossy case.  Here, we therefore ignore the effects of atomic loss.
  
In conclusion, we have solved for the one- and two-photon S-matrices in a waveguide side-coupled to a cavity containing a two-level atom.  Our solution is based on input-output formalism \cite{Gardiner1985}, in Fock-space \cite{Fan2010}, and is fully quantum-mechanical and deterministic.  We have discussed the features of one-photon transport, and the photon-switch effect in both the weak- and strong-coupling regime, as seen in the transport of a two-photon planewave state. Using the S-matrices presented in this paper, the response to one and two incident waveguide photons with arbitrary spectra and pulse shapes may be straightforwardly obtained.

                      \appendix
\section{Derivation of input-output relations for waveguide photon operators}
\label{AppB}
Following the approach of ref. \cite{Fan2010}, we integrate Eq. \eqref{E:ak} from time $ t_0\to -\infty$ to obtain 
\begin{align} a_k(t)=&e^{-\mi k(t-t_0)}a_k(t_0)-\mi V\int_{t_0}^{t} dt'c(t')e^{-\mi k(t-t')}\label{E:AppBakt}
\end{align}
We define $\kappa\equiv 2\pi V^2$, the waveguide-cavity coupling rate.  Additionally,  we define the operator $\Phi(t) \equiv \frac{1}{\sqrt{2\pi}} \int dk a_k(t)$, the input-field operator
 $a_{in}(t)\equiv\frac{1}{\sqrt{2\pi}}\int dke^{-\mi k(t-t_0)}a_k(t_0)$,
  and sum Eq. \eqref{E:AppBakt} over $k$, leading to:
\begin{equation} \Phi(t)=a_{\text{in}}(t)-i\sqrt{\frac{\kappa}{4}}c(t)\label{E:AppBphiIn}
\end{equation}
We also define an output field operator $a_{out}(t)\equiv\frac{1}{\sqrt{2\pi}}\int dke^{-\mi k(t-t_1)}a_k(t_1)$, integrate Eq. \eqref{E:ak} from time $t_1\to \infty$, and sum over $k$, leading to:
\begin{equation}
 \Phi(t)=a_{\text{out}}(t)+\mi\sqrt{\frac{\kappa}{4}}c(t)\label{E:AppBphiOut}
 \end{equation}
Equating Eqs. \eqref{E:AppBphiIn} and \eqref{E:AppBphiOut} we obtain:
\begin{equation}
a_{\text{out}}(t)=a_{\text{in}}(t)-\mi\sqrt{\kappa}c(t).
\label{E:AppBaoutain}
\end{equation}
Plugging  Eq. \eqref{E:AppBphiIn} into  into Eq. \eqref{E:cdot1}, we obtain:
\begin{align}
\frac{dc}{dt}=-\mi\left(\omega-\mi\frac{\kappa}{2}\right)c(t)-\mi\sqrt{\kappa}a_{\text{in}}(t)-\mi g\sigma_-(t).
\label{E:AppBcdot2}
\end{align}
We have thus obtained the input-output formalism [Eqs. \eqref{E:aoutain}, \eqref{E:cdot2}] for this system.
\section{Two-photon planewave transport}
\label{AppC}
Consider an incident two-photon planewave state with right-moving photons:
\begin{align}
\ket{k_1,k_2}_{RR}=\int dx_1dx_2S_{k_1,k_2}(x_1,x_2)\frac{1}{\sqrt{2}}a^{\dag}_R(x_1)a^{\dag}_R(x_2)\ket{0},\label{E:Cplane}\end{align}
where $S_{k_1,k_2}(x_1,x_2)=\frac{1}{\sqrt{2}2\pi}\left[e^{ik_1x_1}e^{ik_2x_2}+e^{ik_1x_2}e^{ik_2x_1}\right]$.  We decompose the state in Eq. \eqref{E:Cplane} into even and odd subspaces by noting that
\begin{align}
&\ket{x_1,x_2}_{RR}=\frac{1}{\sqrt{2}}a^{\dag}_R(x_1)a^{\dag}_R(x_2)\ket{0}
\\&=\frac{1}{\sqrt{2}}\frac{1}{\sqrt{2}}\left[a^{\dag}_e(x_1)+a^{\dag}_o(x_1)\right]\frac{1}{\sqrt{2}}\left[c^{\dag}_{e}(x_2)+c^{\dag}_{o}(x_2)\right]\ket{0}
\\&=\frac{1}{2\sqrt{2}}a^{\dag}_e(x_1)a^{\dag}_e(x_2)\ket{0}+\frac{1}{2\sqrt{2}}a^{\dag}_o(x_1)a^{\dag}_o(x_2)\ket{0}+\frac{1}{2\sqrt{2}}a^{\dag}_e(x_1)a^{\dag}_o(x_2)\ket{0}
\\&+\frac{1}{2\sqrt{2}}a^{\dag}_o(x_1)a^{\dag}_e(x_2)\ket{0}
\\&=\frac{1}{2}\ket{x_1,x_2}_{ee}+\frac{1}{2}\ket{x_1,x_2}_{oo}+\frac{1}{2\sqrt{2}}\ket{x_1,x_2}_{eo}+\frac{1}{2\sqrt{2}}\ket{x_1,x_2}_{oe}.
\end{align}
It follows that
\begin{align}
&\ket{k_1,k_2}_{RR}=\frac{1}{2}\ket{k_1,k_2}_{ee}+\frac{1}{2}\ket{k_1,k_2}_{ee}+\frac{1}{2\sqrt{2}}\ket{k_1,k_2}_{eo}
\\&+\frac{1}{2\sqrt{2}}\ket{k_1,k_2}_{oe},
\end{align}
where $\ket{{k_1,k_2}}_{ee}=\int dx_1dx_2S_{k_1,k_2}(x_1,x_2)\frac{1}{\sqrt{2}}a^{\dag}_e(x_1)a^{\dag}_e(x_2)\ket{0}$ and (e.g) $\ket{k_1,k_2}_{eo}=\int dx_1dx_2S_{k_1,k_2}(x_1,x_2)a^{\dag}_e(x_1)a^{\dag}_o(x_2)\ket{0}$.
We can now apply the scattering operator:
\begin{align}
&\textbf{S}\ket{k_1,k_2}_{RR}=\frac{1}{2}\textbf{S}_{ee}\ket{k_1,k_2}_{ee}+\frac{1}{2}\textbf{S}_{oo}\ket{k_1,k_2}_{oo}+\frac{1}{2\sqrt{2}}\textbf{S}_{eo}\ket{k_1,k_2}_{eo}
\\&+\frac{1}{2\sqrt{2}}\textbf{S}_{oe}\ket{k_1,k_2}_{oe}\label{E:Soperator},
\end{align}
where $\textbf{S}\ket{k_1,k_2}_{RR}$ is the out-state (scattered state).
Below, we work through each one of the terms in Eq. \eqref{E:Soperator}.
\subsection{$eo$ and $oe$ subspaces}
\label{S:eo}
In the $eo$ subspace the scattering operator is $\textbf{S}_{eo}=\textbf{S}_{e}\textbf{S}_{o}=\textbf{S}_{e}\textbf{I}_o$, where $\textbf{I}_o$ is the identity operator in the odd subspace.  We may then rewrite the third term in Eq. \eqref{E:Soperator} \begin{align}
&\frac{1}{2\sqrt{2}}\textbf{S}_{e}\textbf{I}_o\ket{k_1,k_2}_{eo}=
\\&=\frac{1}{2\sqrt{2}}\int dx_1dx_2\frac{1}{\sqrt{2}2\pi}\left[e^{ik_1x_1}e^{ik_2x_2}+e^{ik_1x_2}e^{ik_2x_1}\right]\times
\\&\textbf{S}_{e}a^{\dag}_e(x_1)a^{\dag}_o(x_2)\ket{0}.
\end{align}
Inserting a one-photon resolution of the identity twice, we rewrite as:
\begin{align}
&=\frac{1}{2\sqrt{2}}\int dx_1dx_2\frac{1}{\sqrt{2}2\pi}\left[e^{ik_1x_1}e^{ik_2x_2}+e^{ik_1x_2}e^{ik_2x_1}\right]\times
\\&\int dk\int dp\ket{p}_e\ _e\bra{p} \textbf{S}_{e}\ket{k}_e\  _e\bra{k}x_1\rangle_e \ket{x_2}_o
\end{align}
noting that $_e\bra{p}\textbf{S}_e\ket{k}_e=t_k\delta(k-p)$ we have
\begin{align}
&\frac{1}{2\sqrt{2}}\textbf{S}_{e}\textbf{I}\ket{k_1,k_2}_{eo}
\\&=\frac{1}{2\sqrt{2}}\int dx_1dx_2\frac{1}{\sqrt{2}2\pi}\left[t_{k_1}e^{ik_1x_1}e^{ik_2x_2}+t_{k_2}e^{ik_1x_2}e^{ik_2x_1}\right]\times\\&a^{\dag}_e(x_1)a^{\dag}_o(x_2)\ket{0}.\label{E:eo}
\end{align}
Similarly, in the $oe$ subspace we have:
\begin{align}
&\frac{1}{2\sqrt{2}}\textbf{S}_{e}\hat{\textbf{I}}\ket{k_1,k_2}_{oe}=
\\&=\frac{1}{2\sqrt{2}}\int dx_1dx_2\frac{1}{\sqrt{2}2\pi}\left[t_{k_2}e^{ik_1x_1}e^{ik_2x_2}+t_{k_1}e^{ik_1x_2}e^{ik_2x_1}\right]\times
\\&a^{\dag}_o(x_1)a^{\dag}_e(x_2)\ket{0}\label{E:oe}
\end{align}
\subsection{ $oo$ subspace}
\label{S:oo}
In the $oo$ subspace the scattering operator is $\textbf{S}_{oo}=\textbf{S}_{o}\textbf{S}_{o}=\textbf{I}_{o}\textbf{I}_o$ where $\textbf{I}_o$ is the identity operator in the odd subspace.  Applying the scattering operator we have:
\begin{align}
&\frac{1}{2}\textbf{S}_o\textbf{S}_o\ket{k_1,k_2}_{oo}=\frac{1}{2}\ket{k_1,k_2}_{oo}\end{align}
\subsection{ $ee$ subspace}
\label{S:ee}
In the $ee$ subspace we have
\begin{align}
&\frac{1}{2}\textbf{S}_{ee}\ket{k_1,k_2}_{ee}=\frac{1}{2}\int_{-\infty}^{\infty} dp_1\int_{-\infty}^{p_1}dp_2\ket{p_1,p_2}_{ee}\ _{ee}\bra{p_1,p_2}\textbf{S}_{ee}\ket{k_1,k_2}_{ee},
\end{align}
where we have inserted a two-photon resolution of the identity. We use the result in Eq. \eqref{E:sp1p2k1k2}
\begin{align}
&\frac{1}{2}\textbf{S}_{ee}\ket{k_1,k_2}_{ee}=\frac{1}{2}\int dp_1dp_2 \Bigg\{t_{k_1}t_{k_2}\Bigg[\delta(k_1-p_1)\delta(k_2-p_2)\\&+\delta(k_1-p_2)\delta(k_2-p_1)\Bigg]+s^{(a)}_{p_1}s^{(a)}_{p_2}F(k_1,k_2)\delta(k_1+k_2-p_1-p_2)\Bigg\}\times
\\&\ket{p_1,p_2}_{ee},\label{E:S}
\end{align}
where $\ket{p_1,p_2}_{ee}=\frac{\sqrt{2}}{2\pi}\int dx_1dx_2e^{iE_o x_c}\cos{\Delta_o x}\ket{x_1,x_2}_{ee}$, and $F(k_1,k_2)$ was defined in Eq. \eqref{E:F}.  The first term in Eq. \eqref{E:S}---the uncorrelated transport term---is equal to $\frac{1}{2}t_{k_1}t_{k_2}\ket{k_1,k_2}_{ee}$.  The second term is equal to
\begin{align}
&\frac{\sqrt{2}}{4\pi}F(k_1,k_2)\int dx_1dx_2\ket{x_1,x_2}_{ee}
\\&\int_{-\infty}^{\infty}dE_oe^{iE_o x_c} \delta(E_i-E_o)\int_{-\infty}^{0}d\Delta_o \cos{[\Delta_o x]}s^{(a)}_{p_1}s^{(a)}_{p_2}.
\end{align}
Integrating over $E_o$ and exploiting the invariance of the integrand under the inversion $\Delta_o\to -\Delta_o$:
\begin{align}
&=\frac{\sqrt{2}}{8\pi}F(k_1,k_2)\int dx_1dx_2e^{iE_i x_c}\ket{x_1,x_2}_{ee}\int_{-\infty}^{\infty}d\Delta_o e^{i\Delta_o x}s^{(a)}_{p_1}s^{(a)}_{p_2}
\end{align}
Or:
\begin{align}
&=\frac{\sqrt{2}g^2\kappa}{8\pi}F(k_1,k_2)\int dx_1dx_2e^{iE_i x_c}\ket{x_1,x_2}_{ee}
\\&\int_{-\infty}^{\infty}\frac{d\Delta_o e^{i\Delta_o x}}{\left(\Delta_o+E_i/2-\lambda_{1,-}\right)\left(\Delta_o+E_i/2-\lambda_{1,+}\right)}\times\\&\frac{1}{\left(\Delta_o-E_i/2+\lambda_{1,-}\right)\left(\Delta_o-E_i/2+\lambda_{1,+}\right)}
\end{align}
Integrating over $\Delta_o$ using the residue theorem:
\begin{align}
=&\frac{ig^2\kappa}{2\sqrt{2}}F(k_1,k_2)\int dx_1dx_2e^{iE_i x_c}\times
\\&\Bigg\{e^{i\left(E_i/2-\lambda_{1,-}\right)|x|}\frac{1}{\left(\lambda_{1,+}-\lambda_{1,-}\right)}\frac{1}{\left(E_i-2\lambda_{1,-}\right)\left(E_i-\lambda_{1,-}-\lambda_{1,+}\right)}
\\&+e^{i\left(E_i/2-\lambda_{1,+}\right) |x|}\frac{1}{\left(\lambda_{1,-}-\lambda_{1,+}\right)}\frac{1}{\left(E_i-2\lambda_{1,+}\right)\left(E_i-\lambda_{1,+}-\lambda_{1,-}\right)}\Bigg\}\times\\&\ket{x_1,x_2}_{ee}
\end{align}

Finally, we combine the uncorrelated and correlated terms and get
\begin{align}
&\frac{1}{2}\textbf{S}_{ee}\ket{k_1,k_2}_{ee}=\frac{1}{2}t_{k_1}t_{k_2}\ket{k_1,k_2}_{ee}
\\&+\frac{ig^2\kappa}{2\sqrt{2}}F(k_1,k_2)\int dx_1dx_2e^{iE_i x_c}\times\\&\Bigg\{e^{i\left(E_i/2-\lambda_{1,-}\right)|x|}\frac{1}{\left(\lambda_{1,+}-\lambda_{1,-}\right)}\frac{1}{\left(E_i-2\lambda_{1,-}\right)\left(E_i-\lambda_{1,-}-\lambda_{1,+}\right)}
\\&+e^{i\left(E_i/2-\lambda_{1,+}\right) |x|}\frac{1}{\left(\lambda_{1,-}-\lambda_{1,+}\right)}\frac{1}{\left(E_i-2\lambda_{1,+}\right)\left(E_i-\lambda_{1,+}-\lambda_{1,-}\right)}\Bigg\}
\\&\ket{x_1,x_2}_{ee}.
\end{align}


We then add together the results of sections \ref{S:eo}, \ref{S:oo} and \ref{S:ee}, and use the definition $a^{\dag}_{e}=\frac{a^{\dag}_R(x)+a^{\dag}_L(-x)}{\sqrt{2}}$ and $a^{\dag}_{o}=\frac{a^{\dag}_R(x)-a^{\dag}_L(-x)}{\sqrt{2}}$ to express the out-state in terms of left and right moving photons, as presented in Eqs. \eqref{E:phirr}, \eqref{E:phill} and Eq. \eqref{E:phirl}.

\bibliography{AtomInCavity}

\end{document}